# All Superlinear Inverse Schemes are coNP-Hard[*]

Edith Hemaspaandra[†]
Department of Computer Science
Rochester Institute of Technology
Rochester, NY 14623-5608, USA

Lane A. Hemaspaandra
Department of Computer Science
University of Rochester
Rochester, NY 14627-0226, USA

Harald Hempel[‡]
Institut für Informatik
Friedrich-Schiller-Universität Jena
D-07743 Jena, Germany

October 11, 2004

## Abstract

How hard is it to invert NP-problems? We show that all superlinearly certified inverses of NP problems are coNP-hard. To do so, we develop a novel proof technique that builds diagonalizations against certificates directly into a circuit.

## 1 Introduction

In this paper we show that all superlinear inverse schemes of NP problems are coNP-hard. We develop a novel proof technique that allows us to diagonalize against all possible certificate sets. We feel that this "in-circuit diagonalization" proof technique is of interest in its own right.

The class NP can be viewed as the set of all languages $L$ such that there exist a polynomial-time computable verifier $V$ and a polynomial $q$ such that, for all $x \in \Sigma^*$, $x \in L \iff (\exists y \in \Sigma^*)[|y| = q(|x|) \land V(x,y) \text{ accepts}]$. A string $y$ such that $V(x,y)$ accepts is called a certificate or proof for $x$. Verifiers can formally be defined as follows (see Definition 2.3):

A pair $(V, q)$ is called a standard verifier if and only if

1. $V : \Sigma^* \times \Sigma^* \to \{0, 1\}$ is a polynomial-time computable mapping, and

[*]Supported in part by grants NSF-CCR-9322513, NSF-INT-9815095/DAAD-315-PPP-gü-ab, NSF-CCR-0311021, and NSF-CCF-0426761. Also appears (in a slightly earlier version) as URCS-TR-2004-841.

[†]Work done in part while on sabbatical at the University of Rochester.

[‡]`hempel@informatik.uni-jena.de`. Work done while visiting the University of Rochester.

2. $q : \mathbb{N} \to \mathbb{N}$ is a strictly monotonic, integer-coefficient polynomial such that

$$(\forall x, y \in \Sigma^*)[V(x,y) = 1 \implies |y| = q(|x|)].$$

Inverting standard verification schemes can now informally be described as follows: Let $(V, q)$ be a standard verifier. Given a set $C$ of certificates, does there exist a string $x$ such that $C$ is exactly the set of certificates for $x$ (relative to $(V, q)$)? It is quite natural to choose a succinct representation of certificates, namely, in the form of a circuit. This leads to the following definition (see Definition 2.5) of the inverse problem, which basically asks if a set of strings specified by a circuit is such that some string has precisely those strings as its certificate set.

> Let $(V, q)$ be a standard verifier.
> $\text{Invs}_{V,q} = \{c \mid c \text{ encodes a circuit } c' \text{ having } q(m) \text{ inputs for some } m \in \mathbb{N} \text{ such that } (\exists x \in \Sigma^m)[\{w \in \Sigma^{q(m)} \mid V(x,w) = 1\} = \{y \in \Sigma^{q(m)} \mid c'(y) = 1\}]\}.$

We show that inversion for all superlinear standard verification schemes is coNP-hard. In fact we show even more, namely, that inverting any standard verification scheme $(V, q)$ where $q$ grows faster than all outright linear functions $n + k$, $k \in \mathbb{N}$, is coNP-hard (see Theorem 3.2). So coNP-hardness in fact holds for all $\text{Invs}_{V,q}$ where $(V, q)$ is a standard verification scheme and $q$ is a polynomial of degree either greater than one or of degree one with a degree-one-coefficient $a_1 > 1$.

The proof of our main result is based on a proof technique that can informally be described as an "in-circuit diagonalization" against possible certificate sets. In particular, our in-circuit diagonalization technique uses a circuit *to diagonalize against certificate sets that are potentially accepted by the very same circuit*. The need to diagonalize in such an unusual way arises from the fact that when reducing $\overline{\text{SAT}}$ to $\text{Invs}_{V,q}$ (as we will do in the proof of Theorem 3.2) one has to map boolean formulas to circuits such that the following holds: If the formula is satisfiable then, for all $x$, the set of strings accepted by the circuit is not equal to the set of certificates for $x$ (relative to $(V, q)$); and if the formula is not satisfiable then there exists a string $x$ such that the the set of strings accepted by the circuit is exactly the set of certificates for $x$ (relative to $(V, q)$).

Relatedly, $\Sigma_2^p$ is clearly an upper bound for the complexity of inverting standard verification schemes, and we prove that this upper bound is optimal by constructing a standard verifier such that its inversion problem is $\Sigma_2^p$-complete (see Theorem 3.7). Our actual construction in fact ensures that there exist a P set $A$ and a standard verifier $(V, q)$ for $A$ such that $\text{Invs}_{V,q}$ is $\Sigma_2^p$-complete.

Our results can be extended to also hold for the one-sided variant of inversion of verification schemes, $1\text{-Invs}_{V,q}$. The difference in the definitions of $\text{Invs}_{V,q}$ and $1\text{-Invs}_{V,q}$ (see Definition 2.5) is that instead of requiring "$\exists x \in \Sigma^*$ such that the set of strings accepted by the circuit equals the set of certificates of $x$" as in definition of $\text{Invs}_{V,q}$, we in the definition of $1\text{-Invs}_{V,q}$ require "$\exists x \in L(V, q)$ such that the set of strings accepted by the circuit equals the set of certificates of $x$."

In a fascinating paper by Chen [Che03], a type of inversion of NP problems is studied that is somewhat related to the above-described one-sided-inversion problem, 1-Invs$_{V,q}$, and $\Sigma_2^p$ results are obtained. However, the models are different; for example, in contrast to our definition, where certificates are given in a very succinct form, i.e., implicitly in form of a circuit, Chen studied one-sided inversions of NP problems where the certificates are explicitly given, i.e., in form of a set or a list and, as mentioned above, Chen's focus is on the one-sided inversion problem.

Our paper is organized as follows. After formally defining the basic concepts in Section 2, in Section 3 we state and prove our main result—that all superlinearly certified inverses are coNP-hard. In Section 3 we also prove a number of related theorems, in particular the optimality of the $\Sigma_2^p$ upper complexity bound for Invs$_{V,q}$. In Section 4, we turn to the complexity of recognizing whether machines compute verifiers and we establish $\Sigma_2^0$-completeness results on this.

## 2 Preliminaries

We assume the reader to be familiar with the basic definitions and concepts of complexity theory (see [Pap94, HO02]). Let $\Sigma = \{0,1\}$ be our alphabet. $\mathbb{N}$ denotes $\{0, 1, 2, \ldots\}$ and $\mathbb{Z}$ denotes $\{\ldots, -3, -2, -1, 0, 1, 2, 3, \ldots\}$. We say a polynomial $p$ is strictly monotonic (by which we always mean strictly monotonically increasing) if, for all $n \in \mathbb{N}$, $p(n+1) > p(n)$. For any set $A$ and any $m \in \mathbb{N}$, $A^{=m}$ denotes $\{z \mid z \in A \wedge |z| = m\}$.

Without defining it formally we will make use of a nice (i.e., polynomial-time computable and polynomial-time invertible) encoding of any boolean circuit (consisting of AND, OR and NOT gates) as a word over the alphabet $\Sigma$. As is standard, we denote the outcome (0 or 1, representing reject/false and accept/true) of a circuit $c$ on input $x$ by $c(x)$.

Let FP denote the set of all (total) polynomial-time computable functions, where these functions can be of arbitrary finite arities. We will use the following standard complexity classes.

**Definition 2.1**   1. P *is the set of all languages that can be accepted in deterministic polynomial time.*

2. NP *is the set of all languages that can be accepted in nondeterministic polynomial time.* coNP *is defined to be the set of all languages $\overline{A}$ such that $A \in$ NP.*

3. *[PY84]* DP *is the set of all languages $L$ such that there exist NP sets $A$ and $B$ satisfying $L = A - B$.*

4. *[MS72, Sto76]* $\Sigma_2^p$ *is the set of all languages that can be accepted by nondeterministic polynomial-time Turing machines with the help of an* NP *oracle:* $\Sigma_2^p = \text{NP}^{\text{NP}}$. PH *denotes the polynomial hierarchy:* PH $= \text{P} \cup \text{NP} \cup \text{NP}^{\text{NP}} \cup \text{NP}^{\text{NP}^{\text{NP}}} \cup \cdots$.

We mention in passing that P, NP, and DP are the low levels of the boolean hierarchy [CGH+88, CGH+89] and that P, NP, and $\Sigma_2^p$ are the low levels of the polynomial hierarchy [MS72, Sto76].

Let REC denote the set of all recursive languages. The second level of the arithmetic hierarchy $\Sigma_2^0$ is defined as follows.

**Definition 2.2** *(see [Rog67]) A language $L$ is in $\Sigma_2^0$ if and only if there exists a language $B \in \text{REC}$ such that for all $x \in \Sigma^*$,*

$$x \in L \iff (\exists y \in \Sigma^*)(\forall z \in \Sigma^*)[\langle x, y, z\rangle \in B],$$

*where $\langle \cdot, \cdot, \cdot \rangle$ here is a standard, nice 3-ary pairing function.*

As is standard we will use $\leq_m$ (respectively, $\leq_m^p$) to denote recursive many-one reductions (respectively, polynomial-time many-one reductions) between languages.

In the following we will define the basic concepts that allow us to study inverse NP problems.

**Definition 2.3**   1. *A pair $(V, q)$ is called a* standard verifier *if and only if*

   (a) *$V : \Sigma^* \times \Sigma^* \to \{0, 1\}$ is a polynomial-time computable mapping, and*

   (b) *$q : \mathbb{N} \to \mathbb{N}$ is a strictly monotonic, integer-coefficient polynomial such that*

   $$(\forall x, y \in \Sigma^*)[V(x, y) = 1 \implies |y| = q(|x|)].$$

2. *We say that $(V, q)$ is a standard verifier for a language $L$ if and only if $(V, q)$ is a standard verifier and $L = L(V, q)$, where $L(V, q) = \{x \in \Sigma^* \mid (\exists y \in \Sigma^*)[|y| = q(|x|) \land V(x, y) = 1]\}$ (equivalently, $L(V, q) = \{x \in \Sigma^* \mid (\exists y \in \Sigma^*)[V(x, y) = 1]\}$).*

3. *We say a 2-ary Turing machine $M$ computes a standard verifier if there are a polynomial $r$ and a polynomial $q$ such that*

   (a) *$M$ runs in $r$-bounded time (by which we mean that for each $x, y \in \Sigma^*$, $M(x, y)$ halts in at most $r(|x| + |y|)$ steps), and*

   (b) *$q : \mathbb{N} \to \mathbb{N}$ is a strictly monotonic, integer-coefficient polynomial such that*

   $$(\forall x, y \in \Sigma^*)[\chi_{L(M)}(x, y) = 1 \implies q(|y|) = |x|].$$

   *(Note: Regarding types, $L(M) \subseteq \Sigma^* \times \Sigma^*$, and $\chi_{L(M)}$—the characteristic function—maps from $\Sigma^* \times \Sigma^*$ to $\{0, 1\}$.)*

The following two facts are immediate and standard.

**Fact 2.4**   1. *For every set $A \in \text{NP}$ there exists a standard verifier $(V, q)$ such that $(V, q)$ is a standard verifier for $A$.*

2. If $(V, q)$ is a standard verifier for a language $L$ then $L \in \mathrm{NP}$.

We now define the inverse problem for NP languages.

**Definition 2.5** *Let $A \in \mathrm{NP}$ and let $(V, q)$ be a standard verifier for $A$.*

1. $\mathrm{Invs}_{V,q} = \{c \mid c \text{ encodes a circuit } c' \text{ having } q(m) \text{ inputs for some } m \in \mathbb{N} \text{ such that} \\ (\exists x \in \Sigma^m)[\{w \in \Sigma^{q(m)} \mid V(x,w) = 1\} = \{y \in \Sigma^{q(m)} \mid c'(y) = 1\}]\}.$

2. $1\text{-}\mathrm{Invs}_{V,q} = \{c \mid c \text{ encodes a circuit } c' \text{ having } q(m) \text{ inputs for some } m \in \mathbb{N} \text{ such that} \\ (\exists x \in A^{=m})[\{w \in \Sigma^{q(m)} \mid V(x,w) = 1\} = \{y \in \Sigma^{q(m)} \mid c'(y) = 1\}]\}.$

It is not hard to see that for standard verifiers $(V, q)$, $\mathrm{Invs}_{V,q}$ and $1\text{-}\mathrm{Invs}_{V,q}$ are always in $\Sigma_2^p$. However, $\mathrm{Invs}_{V,q}$ and $1\text{-}\mathrm{Invs}_{V,q}$ seem to differ with respect to their complexity lower bounds.

**Proposition 2.6** *There is a set $A \in \mathrm{NP}$ such that for all standard verifiers $(V, q)$ for $A$, $1\text{-}\mathrm{Invs}_{V,q} \in \mathrm{P}$.*

One proof is by simply choosing $A$ to be $\emptyset$ or any other finite set. In contrast, for every standard verifier $(V, q)$ for $\emptyset$ we have that $\mathrm{Invs}_{V,q}$ is $\leq_m^p$-complete for coNP.

**Proposition 2.7** *Let $(V, q)$ be a standard verifier for $\emptyset$. Then $\mathrm{Invs}_{V,q}$ is $\leq_m^p$-complete for coNP.*

The claim follows from the fact that from $(V, q)$ being a standard verifier for $\emptyset$ the set $\mathrm{Invs}_{V,q}$ is essentially the set of all appropriate-number-of-inputs circuits that for no input evaluate to 1, and is easily seen to be in coNP. Also, it is straightforward to reduce the coNP-complete language $\overline{\mathrm{SAT}}$ to $\mathrm{Invs}_{V,q}$.

## 3  Inverting NP Problems is coNP-complete

Before stating our main theorem we need a technical definition.

**Definition 3.1** *A polynomial $q$ is called* miserly *if and only if for all $\epsilon > 0$ there exist infinitely many $n \in \mathbb{N}$ such that $q(n) \leq (1 + \epsilon)n$.*

Note that for strictly monotonic polynomials $p$, $p(n) = a_k n^k + a_{k-1} n^{k-1} + \cdots + a_1 n + a_0$, with $a_k > 0$, we have that $p$ is nonmiserly if and only if either (a) $k \geq 2$ or (b) $k = 1$ and $a_1 > 1$.

**Theorem 3.2** *Let $A \in \mathrm{NP}$ and $(V, q)$ be a standard verifier for $A$ such that $q$ is a nonmiserly polynomial. Then $\mathrm{Invs}_{V,q}$ is $\leq_m^p$-hard for coNP.*

This immediately yields the following, where by "nonmiserly standard verifier" we mean a standard verifier whose second component is a nonmiserly polynomial.

**Corollary 3.3** *No nonmiserly standard verifier for an* NP *set has an inverse problem belonging to* NP*, unless* NP = coNP*.*

**Proof of Theorem 3.2:** Let $A \in$ NP and let $(V, q)$ be a standard verifier for $A$. Suppose that $q$ is nonmiserly. We will show that $\overline{\text{SAT}} \leq_m^p \text{Invs}_{V,q}$.

Let $F$ be a formula and suppose that $F$ has $n$ variables. Our reduction $g$ will map $F$ to the encoding $c = g(F)$ of a circuit $c'$. The circuit $c'$ will have $q(n')$ inputs where $n'$ is the smallest natural number such that $q(n') > n + n'$. Note that since $q$ is nonmiserly $n'$ is linearly related to $n$ and can be found in polynomial time. On input $z \in \{0,1\}^{q(n')}$, let $x$, $\alpha$, and $r$ be the unique strings such that $z = x\alpha r$, $x \in \{0,1\}^{n'}$, $\alpha \in \{0,1\}^n$, and $r \in \{0,1\}^{q(n')-n'-n}$. The circuit $c'$ consists of three subcircuits that work as follows:

**Subcircuit 1:** Subcircuit 1 simulates the work of $V(x, z)$. Let $a = V(x, z)$ be the output of subcircuit 1.

**Subcircuit 2:** Subcircuit 2 is a polynomial-size-bounded circuit for $F$ with $\alpha$ as its input. Let $b = F(\alpha)$ be the output of subcircuit 2.

**Subcircuit 3:** Subcircuit 3 simulates the work of $V(0^{n'}, z)$. Let $d = V(0^{n'}, z)$ be the output of subcircuit 3.

**Output of $c'$:** $c'$ outputs 0 if $b = d = 0$ or $a = b = 1$. $c'$ outputs 1 otherwise, that is if either a) $b = 0$ and $d = 1$ or b) $b = 1$ and $a = 0$.

It is obvious that $c'$ and thus also $c$ can be constructed in time polynomial in $|F|$.

It remains to show that for all formulas $F$, $F \in \overline{\text{SAT}} \iff g(F) \in \text{Invs}_{V,q}$. Suppose that $F \in \overline{\text{SAT}}$. So we have for all inputs $z$ to the circuit $c'$, $b = 0$. Thus, for all inputs $z$, $c'(z) = 1$ if and only if $d = 1$. By construction $d = 1$ if and only if $V(0^{n'}, z) = 1$. It follows that $\{z \in \Sigma^{q(n')} \mid c'(z) = 1\} = \{y \in \Sigma^{q(n')} \mid V(0^{n'}, y) = 1\}$ and so (via the certificates of $0^{n'}$) $c = g(F) \in \text{Invs}_{V,q}$. For the other direction of the equivalence to be shown assume $F \notin \overline{\text{SAT}}$. So there exists an $n$-bit assignment $\widehat{\alpha}$ for $F$ such that $F(\widehat{\alpha}) = 1$ and consequently for all inputs $z$ to the circuit $c'$ such that $z = x\widehat{\alpha}r$ with $x \in \{0,1\}^{n'}$ and $r \in \{0,1\}^{q(n')-n'-n}$, we have $b = 1$. It follows that for all $x \in \{0,1\}^{n'}$ there exists some input $z'$ to the circuit, namely $z' = x\widehat{\alpha}0^{q(n')-n'-n}$, such that (a) $c'(z') = 0$ if $a = V(x, z') = 1$ and (b) $c'(z') = 1$ if $a = V(x, z') = 0$. It follows that for all $x \in \{0,1\}^{n'}$, $\{z \in \Sigma^{q(n')} \mid c'(z) = 1\} \neq \{y \in \Sigma^{q(n')} \mid V(x, y) = 1\}$, and so $g(F) \notin \text{Invs}_{V,q}$. ❏

Since by our remark preceding Theorem 3.2 any superlinear polynomial is nonmiserly, we have the following corollary.

**Corollary 3.4** *Let $A \in$ NP and let $(V, q)$ be a standard verifier for $A$ such that $q$ is a superlinear polynomial. Then $\text{Invs}_{V,q}$ is $\leq_m^p$-hard for* coNP*.*

Before we can state a similar result for 1-Invs$_{V,q}$, we need a technical concept. Though as far as we know it is a new concept, we feel it is also a very natural concept. We will call this notion P-producibility. (In choosing the nomenclature, we are motivated by the term and notion of "self-P-producible circuits" [Ko85, BB86, GW93].)

**Definition 3.5** *We say a set $A$ is* P-producible *if and only if there exists a function $h \in$ FP, $h : \Sigma^* \to \Sigma^*$, such that for all $x \in \Sigma^*$, $|h(x)| \geq |x|$ and $h(x) \in A$.*

Our definition of P-producibility should be contrasted (especially as to what the polynomial time is in relation to—the input or the output) with the notion of tangibility introduced by Hemachandra and Rudich: A set $A$ is called tangible if and only if there exists a total function $f$ that can be computed in time polynomial in the size of its output such that for all $x \in \Sigma^*$, $f(x) \in A$ and $f(x) \geq_{\text{lexicographical}} x$ [HR90].

**Theorem 3.6** *Let $A$ be any* NP *set that is* P*-producible. Let $(V, q)$ be a standard verifier for $A$ such that $q$ is a nonmiserly polynomial. Then* 1-Invs$_{V,q}$ *is $\leq_m^p$-hard for* coNP.

**Proof:** Let $A$ be an NP set that is P-producible via a function $h \in$ FP, $h : \Sigma^* \to \Sigma^*$. Let $(V, q)$ be a standard verifier for $A$ such that $q$ is a nonmiserly polynomial.

The proof proceeds quite similarly to the proof of Theorem 3.2. Let $F$ be a formula with $n$ variables. Let $n'$ be the smallest natural number such that $q(n') > n + n'$. The difference from the proof of Theorem 3.2 is that the constructed circuit $c'$ has to be modified as follows: Let $w = h(0^{n'+1})$. $c'$ will have $q(|w|)$ inputs. On input $z \in \{0,1\}^{q(|w|)}$, let $z = x\alpha r$ where $x \in \{0,1\}^{|w|}$, $\alpha \in \{0,1\}^n$, and $r \in \{0,1\}^{q(|w|)-|w|-n}$, the circuit works as follows (note the natural adjustment in Subcircuit 3).

**Subcircuit 1:** Subcircuit 1 simulates the work of $V(x, z)$. Let $a = V(x, z)$ be the output of subcircuit 1.

**Subcircuit 2:** Subcircuit 2 is a polynomial-size-bounded circuit for $F$ and uses $\alpha$ as its input. Let $b = F(\alpha)$ be the output of subcircuit 2.

**Subcircuit 3:** Subcircuit 3 simulates the work of $V(w, z)$. Let $d = V(w, z)$ be the output of subcircuit 3.

**Output of $c'$:** $c'$ outputs 0 if $b = d = 0$ or $a = b = 1$. $c'$ outputs 1 otherwise, that is if either a) $b = 0$ and $d = 1$ or b) $b = 1$ and $a = 0$.

The correctness of the reduction can be shown as in the proof of Theorem 3.2, where $w$ now plays the role that $0^{n'}$ played in the proof of Theorem 3.2. ❑

In the reminder of this section we will establish some $\Sigma_2^p$-completeness results and a result about membership in DP.

As already mentioned in Section 2, Invs$_{V,q} \in \Sigma_2^p$ for all standard verifiers $(V, q)$. We will now show that this upper complexity bound is optimal.

**Theorem 3.7** *There exists a standard verifier $(V, q)$ such that $\text{Invs}_{V,q}$ is $\Sigma_2^p$-complete.*

**Proof:** Since $\text{Invs}_{V,q} \in \Sigma_2^p$ for all standard verifiers $(V, q)$, it suffices to show that there exists a standard verifier $(V, q)$ such that $\text{Invs}_{V,q}$ is $\Sigma_2^p$-hard.

Consider the language $\exists\forall 3\text{SAT}$,

$$\exists\forall 3\text{SAT} = \{F \mid F \text{ is a boolean formula in 3-DNF having } 2n \text{ variables}$$
$$x_1, x_2, \ldots, x_n \text{ and } y_1, y_2, \ldots, y_n \text{ for some } n \in \mathbb{N} \text{ and}$$
$$(\exists \alpha \in \{0,1\}^n)(\forall \beta \in \{0,1\}^n)[F(\alpha, \beta) = 1]\},$$

where $F(\alpha, \beta)$ denotes the truth value of $F$ when using $\alpha$ and $\beta$ as assignments for the variables $x_1, x_2, \ldots, x_n$ and $y_1, y_2, \ldots, y_n$, respectively.

$\exists\forall 3\text{SAT}$ is known to be $\Sigma_2^p$-complete [Wra76]. Let *encode* be a polynomial-time computable and polynomial-time invertible encoding function for boolean formulas in 3-DNF. Let *double* be a mapping from $\{0,1\}^*$ to $\{0,1\}^*$ such that for all $k \in \mathbb{N}$ and all $a_1, a_2, \ldots, a_k \in \{0,1\}$, $double(a_1 a_2 \ldots a_k) = a_1 a_1 a_2 a_2 \ldots a_k a_k$.

Let $q(n) = n$ for all $n \in \mathbb{N}$. We define the following verifier $(V, q)$:

> $V$ accepts on input $(u, v)$ if and only if there exist a natural number $n$, a boolean formula $F$ in 3-DNF with $2n$ variables $x_1, x_2, \ldots, x_n, y_1, y_2, \ldots, y_n$, and strings $\alpha, \beta \in \{0,1\}^n$ such that $u = encode(F)01double(\alpha)$ and $v = encode(F)01double(\beta)$ and $F(\alpha, \beta) = 1$.

It is not hard to see that $(V, q)$ is a standard verifier.

To show that $\exists\forall 3\text{SAT} \leq_m^p \text{Invs}_{V,q}$ we will map formulas $F$ having the required syntactic properties (3-DNF, even number of variables) to the encoding $c_F$ of a circuit—having $|encode(F)| + 2n + 2$ inputs—that accepts all strings of the form $encode(F)01double(\beta)$ for any $\beta \in \{0,1\}^n$ and rejects all other strings. All other formulas, i.e., those formulas not in 3-DNF or having an odd number of variables, are mapped to the encoding $c$ of a circuit that accepts exactly one string, namely 0 (this ensures that if $F$ does not have the the required syntactic properties and thus $F \notin \exists\forall 3\text{SAT}$, then $c \notin \text{Invs}_{V,q}$). The described reduction is clearly polynomial-time computable.

It remains to show that for all formulas $F$ having the above-mentioned syntactic properties (3-DNF, even number of variables) it holds that $F \in \exists\forall 3\text{SAT} \iff c_F \in \text{Invs}_{V,q}$. Let $F \in \exists\forall 3\text{SAT}$. It follows that there exists a partial assignment $\alpha \in \{0,1\}^n$ such that for all partial assignments $\beta \in \{0,1\}^n$, $F(\alpha, \beta) = 1$. Hence there exists $u = encode(F)01double(\alpha)$ such that for all $v = encode(F)01double(\beta)$, $V(u, v) = 1$. By construction of $c_F$ we thus have $c_F \in \text{Invs}_{V,q}$. For the other implication assume $F \notin \exists\forall 3\text{SAT}$. Hence for all $\alpha \in \{0,1\}^n$ there exists $\beta \in \{0,1\}^n$ such that $F(\alpha, \beta) = 0$. It follows from the definition of $V$ that for all $u = encode(F)01double(\alpha)$ there exists $v = encode(F)01double(\beta)$ such that $V(u, v) = 0$. By construction of $c_F$ we thus have $c_F \notin \text{Invs}_{V,q}$.

This completes the proof. ❑

Note that the verifier $V$ defined in the proof of Theorem 3.7 is a verifier for the language $L$ of all strings $w$ such that there exist a natural number $n$, a boolean formula $F$ in 3-DNF with $2n$ variables $x_1, x_2, \ldots, x_n, y_1, y_2, \ldots, y_n$, and a string $\alpha \in \{0,1\}^n$ such that

$$w = encode(F)01double(\alpha) \land (\exists \beta \in \{0,1\}^n)[F(\alpha, \beta) = 1].$$

It is not hard to see that $L \in \mathrm{P}$ since satisfiability for 3-DNF formulas can be checked in polynomial time.

**Corollary 3.8** *There exist a language $L \in \mathrm{P}$ and a standard verifier $(V, q)$ for $L$ such that $\mathrm{Invs}_{V,q}$ is $\Sigma_2^p$-complete.*

In fact, looking carefully at the construction in the proof of Theorem 3.7, we see that the just-given proof also establishes the following one-sided result.

**Corollary (to the proof) 3.9** *There exist a language $L \in \mathrm{P}$ and a standard verifier $(V, q)$ for $L$ such that $1\text{-Invs}_{V,q}$ is $\Sigma_2^p$-complete.*

So even simple sets can have very hard inverse problems (Corollaries 3.8 and 3.9). Nonetheless, all (NP) sets have at least one standard verifier whose one-sided inverse problem is not too hard, namely, it belongs to DP (note: if DP = $\Sigma_2^p$ then PH collapses to DP).

**Theorem 3.10** *Every set $A \in \mathrm{NP}$ has a standard verifier $(V, q)$ such that $1\text{-Invs}_{V,q} \in \mathrm{DP}$.*

**Proof:** Let $A \in \mathrm{NP}$ and let $(R, p)$ be a standard verifier for $A$. Let $q(n) = n + p(n)$ and define a verifier $V$ as follows:

> $V$ accepts on input $(a, b)$ if and only if there exists a string $b'$ such that $b = ab'$ and $R(a, b') = 1$.

It is not hard to see that $(V, q)$ is a standard verifier for $A$.

By definition we have

> $1\text{-Invs}_{V,q} = \{c \mid c$ encodes a circuit $c'$ having $q(m)$ inputs for some $m \in \mathbb{N}$ such that $(\exists x \in A^{=m})[\{w \in \Sigma^{q(m)} \mid V(x,w) = 1\} = \{y \in \Sigma^{q(m)} \mid c'(y) = 1\}]\}$.

This can be rewritten, keeping in mind the particular $V$ we have defined, as follows.

$$\text{1-Invs}_{V,q} = \{c \mid c \text{ encodes a circuit } c' \text{ having } q(m) \text{ inputs for some } m \in \mathbb{N}$$

such that:

$(\forall u, v \in \Sigma^{q(m)})[\text{if } c'(u) = c'(v) = 1 \text{ then the first } m \text{ bits of } u \text{ and } v \text{ are identical}]$

and

$(\forall u, v \in \Sigma^{q(m)})(\forall x \in \Sigma^m)[\text{if } c'(u) = 1 \text{ and } c'(v) = 0 \text{ and } u = xu' \text{ and } v = xv' \text{ then } R(x, u') = 1 \text{ and } R(x, v') = 0]$

and

$(\forall u \in \Sigma^{q(m)})(\forall x \in \Sigma^m)[\text{if } c'(u) = 1 \text{ and } u = xu' \text{ then then } R(x, u') = 1]$

and

$(\exists v \in \Sigma^{q(m)})[c'(v) = 1]\}.$

This rewritten version (keeping in mind that the quantification over $m$ is not a "real" quantifier) makes it clear that $\text{1-Invs}_{V,q} \in \text{DP}$, as it is of the form $A \cap B \cap C \cap D$, with $A, B, C \in \text{coNP}$ and $D \in \text{NP}$, and so is of the form of the difference of two NP sets, namely, $D - (\overline{A \cap B \cap C})$. ❑

## 4 The Complexity of Recognizing Verifiers

In this section we show that deciding whether a given machine computes a standard verifier is complete for the second level of the arithmetic hierarchy, $\Sigma_2^0$. Before doing so, we introduce the notion of a "general verifier," in which the "hit the length exactly" restriction on the certificate size is changed to just a one-sided bound, and we prove a $\Sigma_2^0$-completeness result for that. We do so primarily since the proof for that case is clearer and so helps introduce the related but more involved $\Sigma_2^0$-completeness proof for the case of standard verifiers.

**Definition 4.1**  1. A pair $(R, q)$ is called a general verifier *if and only if*

  (a) $R : \Sigma^* \times \Sigma^* \to \{0, 1\}$ *is a polynomial-time computable mapping, and*

  (b) $q : \mathbb{N} \to \mathbb{N}$ *is a strictly monotonic polynomial such that*

  $$(\forall x, y \in \Sigma^*)[R(x, y) = 1 \implies q(|y|) \geq |x|].$$

2. *We say a 2-ary Turing machine $M$ computes a general verifier if there are a polynomial $r$ and a polynomial $q$ such that*

  (a) $M$ *runs in $r$-bounded time (by which we mean that for each $x, y \in \Sigma^*$, $M(x, y)$ halts in at most $r(|x| + |y|)$ steps), and*

  (b) $q : \mathbb{N} \to \mathbb{N}$ *is a strictly monotonic polynomial such that*

  $$(\forall x, y \in \Sigma^*)[\chi_{L(M)}(x, y) = 1 \implies q(|y|) \geq |x|].$$

  *(Note: Regarding types, $L(M) \subseteq \Sigma^* \times \Sigma^*$, and $\chi_{L(M)}$—the characteristic function—maps from $\Sigma^* \times \Sigma^*$ to $\{0, 1\}$.)*

Let $M_1, M_2, M_3, \ldots$ be a standard enumeration of deterministic 2-ary Turing machines.

**Theorem 4.2** *The index set $I_{ver,gen} = \{i \in \mathbb{N} \mid M_i \text{ computes a general verifier}\}$ is $\leq_m$-complete for $\Sigma_2^0$.*

**Proof:** It is not hard to see that $I_{ver,gen} \in \Sigma_2^0$ since $I_{ver,gen}$ can be described as follows: $i \in I_{ver,gen} \iff$

> $(\exists k \in \mathbb{N})(\forall x, y \in \Sigma^*)[M_i(x,y)$ halts within at most $(|x|+|y|)^k + k$ steps and if $M_i(x,y)$ accepts within at most $(|x|+|y|)^k + k$ steps then $|y|^k + k \geq |x|]$.

(To see this, note that given a machine $M_i$ as well as the polynomial $q$ and the strictly monotonic polynomial $r$ that with respect to $M_i$ fulfill part 2 of Definition 4.1, we will choose to use a $k$ so large that $(\forall n \in \mathbb{N})[n^k + k > \max(q(n), r(n))]$.) Note that the right hand side of the above "$\iff$" shows membership in $\Sigma_2^0$.

It remains to show that $I_{ver,gen}$ is $\leq_m$-hard for $\Sigma_2^0$. Since $I_{finite} = \{i \mid L(N_i) \text{ is finite}\}$ (where $N_1, N_2, N_3, \ldots$ is a fixed standard enumeration of Turing machines, e.g., that of Hopcroft–Ullman [HU79]) is $\leq_m$-hard (even $\leq_m$-complete) for $\Sigma_2^0$ it suffices to show that $I_{finite} \leq_m I_{ver,gen}$. Given (as input to our reduction) any $i \in \mathbb{N}$, by the nice properties of the standard enumeration, we can effectively construct from $i$ a machine $E$ that is an enumerator for $L(N_i)$. We now describe a Turing machine $\widehat{M}$. $\widehat{M}$ is a 2-ary Turing machine that on input $(x,y) \in \Sigma^* \times \Sigma^*$ does the following steps:

1. Simulate $|x| + |y|$ steps of the work of $E$ and let $A$ be the set of all strings that are enumerated by $E$ within those $|x| + |y|$ steps.

2. Simulate $2(|x| + |y|)$ steps of the work of $E$ and let $B$ be the set of all strings that are enumerated by $E$ within those $2(|x| + |y|)$ steps.

3. Accept (i.e., output true) if $B - A \neq \emptyset$; otherwise reject (i.e., output false).

Clearly, $\widehat{M}$ is a 2-ary Turing machine. Let $j$ be an index such that $M_j = \widehat{M}$ (we assume our standard enumeration is expansive enough to include all the obviously 2-ary, deterministic machines created by this construction—this is a legal assumption). Since $j$ clearly depends only on $i$ we have implicitly described a mapping $f : \mathbb{N} \to \mathbb{N}$. Note that $f$ is computable.

It suffices to show that for all $i \in \mathbb{N}$, $i \in I_{finite} \iff f(i) \in I_{ver,gen}$. Let $i \in \mathbb{N}$ and let $j = f(i)$.

*Case 1:* $i \in I_{finite}$. So $L(N_i)$ is finite and the number of strings enumerated by $E$ is finite as well. Note that since $M_j$ by definition runs in polynomial time and since $E$ enumerates only a finite number of strings it follows from the construction of $M_j$ that $M_j$ accepts only a finite number of inputs and thus it holds that there exists a strictly monotonic (integer-coefficient) polynomial $p$ such that for all $x, y \in \Sigma^*$, if $M_j(x,y)$ outputs true then $p(|y|) \geq |x|$. So (remembering also the polynomial-time claim made above) $M_j$ computes a general verifier and thus $j \in I_{ver,gen}$.

*Case 2: $i \notin I_{finite}$.* In this case, $E$ enumerates an infinite number of strings and thus for all $y \in \Sigma^*$, $M_j(x,y)$ outputs true for infinitely many $x \in \Sigma^*$. So there does not exist a (strictly monotonic) polynomial $p$ such that, for all $x, y \in \Sigma^*$, if $M_j(x,y)$ outputs true then $p(|y|) \geq |x|$. Thus, $M_j$ does not compute a general verifier and so $j \notin I_{ver,gen}$. ❑

Does the same classification hold for standard verifiers? Note that the "hit the length on the head"-ness of standard verifiers will be something of a technical obstacle. Nonetheless, by carefully choosing the pairs $(x,y)$ that are accepted by the constructed machine we are able to show that deciding whether a given machine computes a standard verifier is also complete for $\Sigma_2^0$.

**Theorem 4.3** *The index set $I_{ver,std} = \{i \in \mathbb{N} \mid M_i \text{ computes a standard verifier}\}$ is $\leq_m$-complete for $\Sigma_2^0$.*

**Proof:** It is not hard to see that $I_{ver,std} \in \Sigma_2^0$ since $I_{ver,std}$ can be described as follows: $i \in I_{ver,std} \iff$

$(\exists k \in \mathbb{N})(\exists \ell \in \mathbb{N})(\exists a_0, a_1, a_2, \ldots a_\ell \in \mathbb{Z})(\forall x, y \in \Sigma^*)(\forall n \in \mathbb{N})[(M_i(x,y)$ halts within at most $(|x| + |y|)^k + k$ steps and if $M_i(x,y)$ accepts within at most $(|x|+|y|)^k + k$ steps then $|y| = a_\ell |x|^\ell + a_{\ell-1}|x|^{\ell-1} + \cdots + a_1 |x| + a_0])$ and $(a_\ell n^\ell + a_{\ell-1} n^{\ell-1} + \cdots + a_1 n + a_0 < a_\ell(n+1)^\ell + a_{\ell-1}(n+1)^{\ell-1} + \cdots + a_1(n+1) + a_0)]$.

Note that the right hand side of the above " $\iff$ " shows membership in $\Sigma_2^0$.

It remains to show that $I_{ver,std}$ is $\leq_m$-hard for $\Sigma_2^0$. As in the proof of Theorem 4.2, it suffices to show that $I_{finite} \leq_m I_{ver,std}$.

Before we describe the reduction we need a few technical definitions. We define a family of polynomials as follows:
$$q_0(n) = n + 1,$$
and, for each $i \in \{1, 2, 3, \ldots\}$, we inductively define
$$q_{i+1}(n) = n(n-1)(n-2)\ldots(n-i) + q_i(n).$$

Note that, for all $i \in \mathbb{N}$, (a) $i! \leq q_i(i)$ and (b) $q_i$ is a strictly monotonic, integer-coefficient polynomial. Observe that the polynomials $q_i$ and the numbers $m_i$ satisfy the following:

$$
\begin{array}{llllllllllll}
1 & = & q_0(0) & = & q_1(0) & = & q_2(0) & = & q_3(0) & = & q_4(0) & = & q_5(0) & = & \ldots \\
3 & = & & & q_1(1) & = & q_2(1) & = & q_3(1) & = & q_4(1) & = & q_5(1) & = & \ldots \\
7 & = & & & & & q_2(2) & = & q_3(2) & = & q_4(2) & = & q_5(2) & = & \ldots \\
\vdots & & & & & & & \ddots & & & & & & & \\
m_i & = & & & & & & & q_i(i) & = & q_{i+1}(i) & = & q_{i+2}(i) & = & \ldots \\
\vdots & & & & & & & & & \ddots & & & & &
\end{array}
$$

We return to showing that $I_{finite} \leq_m I_{ver,std}$. So, suppose that we are given any $i \in \mathbb{N}$ (and we wish to effectively compute a string $f(i)$ such that $i \in I_{finite} \iff f(i) \in I_{ver,std}$). By the nice properties of the standard enumeration, we can effectively construct from $i$ a machine $E$ that is an enumerator for $L(N_i)$. We now describe a Turing machine $\widehat{M}$. $\widehat{M}$ is a 2-ary Turing machine that on input $(x,y) \in \Sigma^* \times \Sigma^*$ does the following steps:

1. If $|y| \neq q_{|x|}(|x|)$ halt and reject the input (i.e., output false). If $|y| = q_{|x|}(|x|)$ continue.

2. Simulate $|x| + |y|$ steps of the work of $E$ and let $A$ be the set of all strings that are enumerated by $E$ within those $|x| + |y|$ steps.

3. Simulate $(|x| + |y|)^2$ steps of the work of $E$ and let $B$ be the set of all strings that are enumerated by $E$ within those $(|x| + |y|)^2$ steps.[1]

4. Accept (i.e., output true) if $B - A \neq \emptyset$; otherwise reject (i.e., output false).

Clearly, $\widehat{M}$ is a 2-ary Turing machine. Let $j$ be an index such that $M_j = \widehat{M}$ (we assume our standard enumeration is expansive enough to include all the obviously 2-ary, deterministic machines created by this construction—this is a legal assumption). Since $j$ clearly depends only on $i$ we have implicitly described a mapping $f : \mathbb{N} \to \mathbb{N}$. Note that $f$ is computable.

It suffices to show that for all $i \in \mathbb{N}$, $i \in I_{finite} \iff f(i) \in I_{ver,std}$. Let $i \in \mathbb{N}$ and let $j = f(i)$.

*Case 1:* $i \in I_{finite}$. So $L(N_i)$ is finite and the number of strings enumerated by $E$ is finite as well. Note that since $M_j$ by definition runs in polynomial time and since $E$ enumerates only a finite number of strings it follows from the construction of $M_j$ that $M_j$ accepts only a finite number of inputs. So it holds that there exists a strictly monotonic (integer-coefficient) polynomial $p$ such that for all $x, y \in \Sigma^*$, if $M_j(x, y)$ outputs true then $p(|x|) = |y|$. In particular, by our definition of the polynomials $q_i$ (and remembering also the polynomial-time claim made above) we have that if $\widehat{n} \in \mathbb{N}$ is the largest number such that a pair $(x, y)$, $|x| = \widehat{n}$, is accepted by $M_j$ then ($|y| = q_{\widehat{n}}(\widehat{n})$ and) $M_j$ computes a standard verifier (with $q_{\widehat{n}}$ working as the "$q$" of Part 3 of Definition 2.3).

*Case 2:* $i \notin I_{finite}$. In this case, $E$ enumerates an infinite number of strings. We will argue that then $M_j$ accepts an infinite number of pairs and thus there does not exist a polynomial $p$ such that for all pairs $(x, y) \in L(M_j)$ we have $|y| = p(|x|)$. Note that the Turing machine $M_j$ described above accepts only pairs $(x, y)$ where $|y| = q_{|x|}(|x|)$ and thus one might worry that even though $E$ enumerates an infinite set, $M_j$ only accepts finitely many pairs. Indeed, observe that if we had (as in the proof of Theorem 4.2) chosen the number of steps $E$ is simulated in steps 2 and 3 of the description of $M_j$ to be, respectively, $|x| + |y|$ and $2(|x| + |y|)$, we would have left coverage "gaps," and it might happen that even though $E$ enumerates an infinite set, $M_j$ would be "triggered" to accept pairs only a finite number of times. However, by choosing the number of steps the enumerator $E$ is simulated by $M_j$ to be $|x| + |y|$ and $(|x| + |y|)^2$ in, respectively, steps 2 and 3, it follows[2] that if $E$ enumerates an infinite set then $M_j$ accepts infinitely many pairs. So $M_j(0^n, 0^{q_n(n)})$, when $i \notin I_{finite}$, outputs true for infinitely many $n \in \mathbb{N}$. Recall that by definition we have that for all $n \in \mathbb{N}$, $n! \leq q_n(n)$. So there does not exist a polynomial $p$ (whether strictly monotonic or otherwise) such that, for all $x, y \in \Sigma^*$, if $M_j(x, y)$ outputs true then $p(|x|) = |y|$. Thus, $M_j$ does not compute a standard verifier, and so $j \notin I_{ver,std}$. ❑

---

[1] The reason that we use the bound $(|x|+|y|)^2$, rather than $2(|x|+|y|)$ as we did in the proof of Theorem 4.2, will be explained later in this proof.

[2] Keeping in mind that the only interesting case is when the second argument's length, call it $\ell_2$, is related to the first argument's length, call it $\ell_1$, by the equation $\ell_2 = q_{\ell_1}(\ell_1)$, what we need to show to ensure that there are only finitely many gaps in coverage is that for all but at most a finite number of $n$'s (and only focusing in this footnote on second arguments $y$ of the length-relation just mentioned) the simulation-step bound in Step 3 when $|x| = n$ is greater than the simulation-step bound in Step 2 when $|x| = n + 1$. That is, we need it to hold that, for all sufficiently large $n \in \mathbb{N}$, $(n + q_n(n))^2 \geq (n + 1) + q_{n+1}(n + 1)$.

## 5 Conclusions

We have shown that all superlinear inversion schemes are coNP-hard. We have also shown that some inversion schemes are $\Sigma_2^p$-complete. Note that for finite sets $A$ and any of their standard verifiers $(V, q)$ we have that $\text{Invs}_{V,q}$ is coNP-complete. It is not clear whether the complexity of inverting standard verifiers for infinite NP sets is also independent of the verifier. In particular, does every infinite NP set have a standard verifier $(V, q)$ such that $\text{Invs}_{V,q}$ is $\Sigma_2^p$-complete?

---

Regarding the right-hand side, note that we can inductively see from the definition of the $q_i$'s that, for all $n \in \mathbb{N}$, $q_{n+1}(n+1) \leq ((n+1)!)(n+2)$. And by the lower bound given earlier for $q_n(n)$, we know also that, for all $n \in \mathbb{N}$, $(n + q_n(n))^2 \geq (n + (n!))^2 = n^2 + 2n(n!) + (n!)^2$. We thus are done, since clearly for all sufficiently large $n \in \mathbb{N}$ it holds that (note the asymptotically very large $(n!)^2$ term of the former) $n^2 + 2n(n!) + (n!)^2 \geq (n+1) + ((n+1)!)(n+2)$.